\documentclass[10pt]{iopart}

\usepackage{iopams}
\usepackage{setstack}

\newcommand{\galg}{\mathfrak{g}}
\newcommand{\bc}{\mathrm{b}^\dagger}
\newcommand{\ba}{\mathrm{b}}
\newcommand{\fc}{\mathrm{c}^\dagger}
\newcommand{\fa}{\mathrm{c}}
\newcommand{\Ham}{\mathcal{H}}
\newcommand{\average}[1]{\langle {#1} \rangle}
\newcommand{\bra}[1]{\langle {#1} \! \! \mid}
\newcommand{\ket}[1]{\mid \! \! {#1} \rangle}
\newcommand{\mods}[1]{\mid \! {#1} \! \mid}
\newcommand{\braket}[2]{\langle {#1} \! \! \mid \! \! {#2} \rangle}
\newcommand{\comm}[2]{\left[ {#1},{#2} \right]}
\newcommand{\acom}[2]{\left\{ {#1},{#2} \right\}}
\newcommand{\vect}[2]{\left( \begin{array}{c} {#1} \\ {#2} \end{array} \right)}

\newcommand{\matr}[4]{\left( \begin{array}{cc} {#1}& {#2} \\ {#3} & {#4} \end{array} \right)}
\newcommand{\Lop}{\mathrm{L}}

\begin{document}
\title[A minimal approach for the local statistical properties of a disordered wire]{A minimal approach for the local statistical properties of a one-dimensional disordered wire}
\author{M. Ancliff and B.A. Muzykantskii}
\address{Department of Physics, University of Warwick, Coventry CV4 7AL, England}

\begin{abstract}
We consider a one-dimensional wire in gaussian random potential. By treating the spatial direction as imaginary time, we construct a `minimal' zero-dimensional quantum system such that the local statistical properties of the wire are given as products of statistically independent matrix elements of the evolution operator of the system. The space of states of this quantum system is found to be a particular non-unitary, infinite dimensional representation of the pseudo-unitary group, $U(1,1)$. 

We show that our construction is minimal in a well defined sense, and compare it to the supersymmetry and Berezinskii techniques.
\end{abstract}


\maketitle

\section{Introduction}
The problem of finding statistical properties of disordered one-dimensional wires has been successfully treated by a number of different techniques (see \cite{Beenakker} and references therein, also \cite{Abrikosov,Berezinskii,BF}). By interpreting the x-coordinate as imaginary time, the problem can be reduced to the study of an ensemble of time-dependent Hamiltonians acting on some space of states. At each point in time, in each realisation of disorder, the Hamiltonian operator lies in a Lie algebra of operators on the space - the so-called `dynamical algebra'. In this paper we aim to develop this idea into a framework in which models are defined in terms of a dynamical algebra, and its action on a given linear space. In the supersymmetric treatment one obtains a supersymmetric Fock space and a Lie super-algebra \cite{BF,Bocquet}. Other techniques for the one-dimensional wire do not explicitly involve the action of such a dynamical algebra on a space of states, though such structure can be seen implicitly. The Berezinskii technique \cite{Berezinskii} is one such example, and we discuss it in relation to the framework we develop.

Within this framework we define a notion of minimality in terms of the dynamical algebra, the space on which it acts and a set of physical quantities to be studied. We construct such a minimal description which contains complete information on the local statistical properties of the wire. The choice of physical properties is important - by restricting ourselves to local properties only we are able to construct a simpler model than would be possible if we were to insist on a full knowledge of all the statistical properties. The supersymmetric treatment\cite{Bocquet,BM} is not minimal for the local statistical properties in the sense we define, nor indeed for any larger set. It can be shown that the space we construct is isomorphic to a subspace of the supersymmetric Fock space, though our construction is direct and does not depend upon this.

We hope that the construction we present is generic enough to be applicable to other disordered systems which are not so well understood. In more complicated systems where a localisation-delocalisation transition occurs knowledge of the local properties is enough to determine which state the system is in. The hope is that through focussing on the simplest set of physical characteristics and by adapting the approach developed in this paper, it may be possible to obtain a construction which can more readily be analysed than those which presently exist.

\section{The minimal construction for the local properties}\label{main}
\subsection{Outline of the framework}

In the general theory of disordered systems, one begins with an ensemble of disorder realisations $V$ in $d$-dimensional space, with a given probability distribution. One way to analyse such a system is to interpret one spatial direction as imaginary time, $\tau$, and thereby construct an associated $(d-1)$-dimensional quantum system. In this case each particular realisation of disorder corresponds to a time-dependent Hamiltonian, $V \leftrightarrow H_V(\tau)$ acting on the space of states of the quantum system. Physical quantities are given by expressions involving matrix elements of the evolution operator, $U_V$, in this space.

An additional structure is the `dynamical algebra', $\galg$ which acts on the space of states, and is the linear closure (with respect to commutators) of all possible Hamiltonian operators, at each point in time, in each disorder realisation. We use the representation theory of $\galg$ as a powerful tool to simplify our analysis.

For example, for the ($d=1$) disordered wire, we study the Dirac Hamiltonian in the presence of gaussian disorder,
\begin{equation}\label{Hamiltonian}
\Ham = -\rmi\sigma_3 \frac{\rmd}{\rmd x} + V(x)
\end{equation}
where $V$ assigns a $2\times 2$ Hermitian matrix to each point,
\begin{equation}
V(x) = \matr{\alpha}{z}{z^\ast}{\beta} \quad \alpha,\beta \in \mathbb{R}\ ,\ z \in \mathbb{C}
\end{equation}

Interpreting $x$ as imaginary time $\tau$, the equation for $U_V$ has the form
\begin{equation}\label{quantHam}
\frac{\rmd U_V}{\rmd \tau}(\tau) = -\rmi\sigma_3 V(\tau)U_V(\tau) \ :\ U_V(\tau_0)=1
\end{equation}
So we can view $H_V(\tau) = -\rmi\sigma_3 V(\tau)$ as a zero-dimensional Hamiltonian. Since $V(\tau)$ is Hermitian, $\rmi\sigma_3 V(\tau)$ lies in the $u(1,1)$ algebra, which is the dynamical algebra for this system. In this case $U_V$ is nothing but the transfer matrix for the original Hamiltonian.

Next we need to construct a space on which the dynamical algebra acts. The choice of space will depend upon what physical quantities are to be studied, according to the requirement that in an appropriate space these quantities will be given by products of statistically independent matrix elements of $U_V$ in this space. The statistical independence of these expressions will be important when calculating the averages of the physical quantities over disorder realisations. This is illustrated by the following example for the one-dimensional wire.

In order to perform the averaging we introduce a probability measure on the space of disorder matrices,
\setlength\arraycolsep{2pt}
\begin{equation}\label{ProbDist}
P(V(x)) = A \exp \left[-2l_f(\alpha^2 + \beta^2) - 2l_b\mods{z}^2\right]
\end{equation}
\setlength\arraycolsep{5pt}
and assume the disorder to be spatially delta-correlated. Here $l_f$ and $l_b$ are the forward and backward scattering lengths, respectively.

We will be interested in the reflection amplitudes for a particle in the wire. The amplitude of reflection for a left-moving particle, $r_L$, can be written in terms of the components of the matrix $U_V$:
\begin{equation}\label{TLRA}
r_L(\tau) = \frac{\left( U_V(\tau)\right)_{12}}{\left(U_V(\tau)\right)_{22}}
\end{equation}
The natural space on which $H_V$ acts is the space of two-component complex vectors (the spin $1/2$ representation). From (\ref{TLRA}) the left reflection amplitude is a ratio of two matrix elements of $U_V$ on this space.

Averaging $U_V$ over disorder realisations to obtain an averaged evolution operator, $U_{av}$, is straightforward. However, when working in the natural (spin-$1/2$) representation, averaging the expression for $r_L$ is still non-trivial\footnote{Of course, $r_L$ averages to zero due to uniform averaging over its phase, here it is simply used as an illustration. Later we will consider moments of the square modulus of $r_L$, which are non-trivial.}, as it is the ratio of two statistically \emph{dependent} matrix elements of $U_V$. Rather than using the natural representation, we therefore look for an associated representation of the dynamical algebra, where the expression for $r_L$ in terms of matrix elements of $U_V$ involves only statistically independent products, and hence its averaged value can be written directly in terms of matrix elements of $U_{av}$ - such a representation is constucted in the next subsection.

The analysis for the disordered wire above motivates a general framework for the treatment of disordered systems, given by a dynamical algebra which contains all possible disorder Hamiltonians, a set of physical quantities to be studied, and a representation of the dynamical algebra which is appropriate for this set of quantities, in the sense defined above. We call such a construction \emph{minimal} with respect to the set of physical quantities if there is no strict sub-representation which is also appropriate for this set. In our treatment (section \ref{construct}), we construct a representation of the dynamical algebra directly from the requirement that it is minimal with respect to the local properties of the wire.

\subsection{Construction for the left reflection amplitude}

We use the following notation for a basis of (the complexification of) the $u(1,1)$ algebra:
\setlength\arraycolsep{2pt}
\begin{equation}
\Lop_0 = \frac{1}{2}\matr{1}{0}{0}{-1} \quad \Lop_+ = \matr{0}{1}{0}{0} \quad \Lop_- = \matr{0}{0}{1}{0}
\end{equation}
along with the central element, I. Since $u(1,1)$ is the dynamical algebra for the disordered wire, in each realisation of disorder the evolution operator $U_V$ will lie in the $U(1,1)$ group.

To extract the left reflection amplitude from the evolution operator, we note the relation:
\begin{equation}
U_V(\tau,-\infty) \vect{0}{t_L} = \vect{r_L}{1}
\end{equation}
where $r_L$, $t_L$ are the amplitudes of reflection and transmission for a left-moving particle at position $\tau$. Hence the evolution operator can be written,
\begin{equation}\label{Tleft}
U_V(\tau,-\infty) = \matr{\rme^{\rmi\phi_L}/t^\ast_L}{r_L/t_L}{r^\ast_L \rme^{\rmi\phi_L}/t^\ast_L}{1/t_L}
\end{equation}
where $\phi_L$ is some phase undetermined by the above relation.

The following triangular decomposition of $U_V(\tau,-\infty)$ in terms of algebra elements is useful:
\begin{equation} \label{decomposition}
U_V(\tau,-\infty) = \rme^{2\rmi\theta_1 \Lop_0} \rme^{{\scriptscriptstyle R} \Lop_+}\rme^{2 \ln {\scriptscriptstyle T} \Lop_0}
\rme^{{\scriptscriptstyle R} \Lop_-}\rme^{2\rmi\theta_2 \Lop_0}\rme^{\rmi\phi_L (\Lop_0 + \mathrm{I}/2)}
\end{equation}
where $r_L={\scriptstyle R}\rme^{2\rmi\theta_1}$ and $t_L={\scriptstyle T} \rme^{\rmi(\theta_1+\theta_2)}$ for real $\scriptstyle R$, ${\scriptstyle T}$, $\theta_1$ and $\theta_2$.
In order to simplify the notation, when we consider the action of the $U(1,1)$ group (or algebra) in a given representation we shall not explicitly denote the mapping from the group (or algebra) to the space of linear operators acting on the representation. Although we will work with a number of different representations of $U(1,1)$ in what follows, the implied action of $U(1,1)$ will be made clear by the context.

We now construct a representation, $T$, of $U(1,1)$, such that the left reflection amplitude can be written as a matrix element of the evolution operator in this representation:
\begin{equation}\label{rL}
r_L = \bra{\varphi} U_V(\tau,-\infty) \ket{\chi} \quad :\quad \ket{\chi}\in T, \bra{\varphi}\in T^\ast
\end{equation}

Consider the action of $U_V$ in the representation $T$. The matrix element (\ref{rL}) must be independent of ${\scriptstyle T}$, $\theta_2$, and $\phi_L$, and so we chose a state $\ket{\chi}$ such that the first (rightmost) of the four terms in the decomposition (\ref{decomposition}) act trivially on it:
\begin{equation}\label{Vconds}
\Lop_0 \ket{\chi} = \Lop_- \ket{\chi} = \mathrm{I} \ket{\chi} = 0
\end{equation}

Consider the representation generated by applying algebra elements to this state. The condition $\Lop_+ \ket{\chi} \neq 0$ is enough to specify the representation uniquely (if $\Lop_+ \ket{\chi} = 0$ the trivial representation is obtained). From (\ref{Vconds}), $\ket{\chi}$ is a lowest weight vector with $\Lop_0$-weight zero. Henceforth we denote it by $\ket{0^-}$, and the representation it generates by $T^-_0$. The representation is non-unitary and has infinite dimension (see \cite{Vilenkin}, section 6.4). It is spanned by the vectors $\ket{n^-}=(\Lop_+)^n \ket{0^-}$, with $\Lop_0$ weights $n \geq 0$.

From (\ref{decomposition}) we have,
\begin{eqnarray}
U_V(\tau,-\infty)\ket{0^-} & = & \rme^{2\rmi\theta_1 \Lop_0} \rme^{\scriptstyle R \Lop_+} \ket{0^-} \\
	& = & \sum_{n=0}^{\infty} \frac{(r_L)^n}{n!} \ket{n^-} \label{rLn}
\end{eqnarray}
Hence by setting $\bra{n^-}\in T^\ast$ such that $\braket{n^-}{m^-}=n!\delta_{nm}$, and taking $\bra{\varphi}=\bra{1^-}$, we arrive at the formula for $r_L$ as a matrix element of $U_V$ in this representation:
\begin{equation}
r_L = \bra{1^-}U_V(\tau,-\infty)\ket{0^-}
\end{equation}

Note that there are similar expressions for arbitrary powers of $r_L$ -- from equation (\ref{rLn}) we have,
\begin{equation}\label{LRAs}
{r_L}^n = \bra{n^-}U_V(\tau,-\infty)\ket{0^-}
\end{equation}
This completes our construction for the left-reflection coefficient, raised to any power.

\subsection{Construction for the local statistical properties}\label{construct}

Expressions for the left-reflection coefficient alone are not sufficient to calculate all the local properties of the wire. However, the local Greens functions for a one-dimensional wire at point $\tau$ can be written in terms of the reflection amplitudes for a left-moving and right-moving electron at that point \cite{STBB},
\begin{equation}\label{Greens}
G^R(\tau) = \frac{(1+r_L(\tau))(1+r_R(\tau))}{\rmi(1-r_R(\tau)r_L(\tau))}
\end{equation}
By expanding this expression in positive powers of the reflection amplitudes, it follows that all local properties of the system (and their moments) can be written as infinite sums of terms of the form ${r_L}^n {r_L^\ast}^{\bar{n}}{r_R}^m {r_R^\ast}^{\bar{m}}$, for $n,\bar{n},m,\bar{m}\in \mathbb{N}_0$.

The representation we constructed in the previous section gives expressions for $r_L^n$ in terms of matrix elements of $U_V$. We require similar expressions for the complex conjugate, ${r^\ast_L}^n$. To find the appropriate representation we make use of the algebra automorphism:
\begin{equation}
\Lop_0 \to -\Lop_0 \ ,\  \Lop_+ \to \Lop_- \ ,\ \Lop_- \to \Lop_+
\end{equation}
In the natural representation of $u(1,1)$ this automorphism has the effect of swapping the channel indices, $1 \leftrightarrow 2$. From equations (\ref{TLRA}) and (\ref{Tleft}), this maps $r_L$ into $r_L^\ast$:
\begin{equation}
r_L = \frac{\left( U_V\right)_{12}}{\left(U_V\right)_{22}} \longmapsto
\frac{\left( U_V\right)_{21}}{\left(U_V\right)_{11}} = r_L^\ast
\end{equation}

Using this algebra automorphism, we obtain from $T^-_0$ a new representation, which we denote $T^+_0$. This representation is generated by highest weight vector $\ket{0^+}$, such that
\begin{displaymath}
\Lop_0 \ket{0^+} = \Lop_+ \ket{0^+} = \mathrm{I} \ket{0^+} = 0\ ,\ \Lop_- \ket{0^+} \neq 0
\end{displaymath}
Similarly to the above, we set $\ket{n^+} = (\Lop_-)^n \ket{0^+}$, $\braket{n^+}{m^+}=n!\delta_{nm}$, and obtain the result:
\begin{equation}\label{CLRAs}
(r^{\ast}_L)^n = \bra{n^+}U_V(\tau,-\infty)\ket{0^+}
\end{equation}
It follows that any product ${r_L}^n {r_L^\ast}^{\bar{n}}$ can be expressed as a matrix element of the evolution operator in the tensor-product representation, $T^-_0\otimes T^+_0$:
\begin{equation}
{r_L}^n {r_L^\ast}^{\bar{n}} = \bra{n^-\otimes\bar{n}^+}\ U_V(\tau,-\infty)\ \ket{0^-\otimes 0^+}
\end{equation}

To get an analogous expression for the right reflection amplitude, we note that the transformation $x \to -x$ combined with swapping the order of the channels in the wire, swaps the right and left reflection amplitudes. Hence,
\begin{equation}\label{RRAs}
(r_R)^n = \bra{n^-} \left( \sigma_x U_V(\tau,\infty) \sigma_x \right) \ket{0^-}
\end{equation}
\begin{equation}\label{CRRAs}
(r^\ast_R)^n = \bra{n^+} \left( \sigma_x U_V(\tau,\infty) \sigma_x \right) \ket{0^+}
\end{equation}
So powers of the right reflection amplitude can be obtained from the same representations $T^-_0$ and $T^+_0$.

We are now in a position to write an arbitrary product of reflection coefficients in terms of matrix elements of $U_V$ in the representation $T^-_0 \otimes T^+_0$:
\begin{eqnarray}
{r_L}^n {r_L^\ast}^{\bar{n}}{r_R}^m {r_R^\ast}^{\bar{m}} & = &
\left(\bra{n^-\otimes\bar{n}^+}U_V(\tau,-\infty)\ket{0^-\otimes 0^+}\right) \times \nonumber \\
&& \quad \times \left(\bra{m^-\otimes\bar{m}^+}
\left( \sigma_x U_V(\tau,\infty) \sigma_x \right)\ket{0^-\otimes 0^+}\right) \label{moments}
\end{eqnarray}
Since the intervals $(-\infty,\tau)$ and $(\tau,\infty)$ are disjoint and the disorder is delta correlated, the individual matrix elements are statistically independent and can be averaged seperately.

As was noted above, all local properties of the system (and their moments) can be written as infinite sums of these expressions, which means that the representation $T^-_0 \otimes T^+_0$ is appropriate for the set of local properties. In fact, from the form of the Greens function (\ref{Greens}) it follows that for any $n,\bar{n},m,\bar{m}$, the term ${r_L}^n {r_L^\ast}^{\bar{n}}{r_R}^m {r_R^\ast}^{\bar{m}}$ will occur in the expansion of some product of retarded and advanced Greens functions. Since the states $\ket{n^-}\otimes\ket{\bar{n}^+}$ span the whole of $T^-_0 \otimes T^+_0$, this implies that the space we have constructed is minimal for the local statistical properties of the wire, in the sense defined above.

\subsection{Averaging}

Now that we have constructed the space $T^-_0 \otimes T^+_0$, we show how to average the evolution operator $U_V$ on this space. Averaging equation (\ref{quantHam}) with the probability distribution given in (\ref{ProbDist}), we obtain an equation for the averaged evolution operator, $U_{av}$,
\begin{equation}\label{UavEvolution}
\frac{\rmd U_{av}}{\rmd \tau}=\frac{1}{2}\average{(-\rmi\sigma_3 V)^2}U_{av}
\end{equation}
\begin{equation}\label{Hav}
\frac{1}{2}\average{(-\rmi\sigma_3 V)^2} = -\frac{1}{4l_f}({\Lop_0}^2 + \mathrm{I}^2/4) + \frac{1}{4l_b}(\Lop_+\Lop_- + \Lop_-\Lop_+) =: H_{av}
\end{equation}
The triangular brackets denote averaging\footnote{It is important to remember that $\rmi\sigma_3 V$ should be treated as an operator on $T^-_0 \otimes T^+_0$, so $(\rmi\sigma_3 V)^2$ is the result of applying this operator twice -- it is not the square of the matrix in the natural representation.}. Moments of reflection amplitudes are given by matrix elements of $U_{av}$, in the form of equation (\ref{moments}).

Under the assumption of strong forward scattering $l_f \to 0$, all eigenstates of $\Lop_0$ with non-zero eigenvalue are killed by $U_{av}$, because of the presence of ${l_f}^{-1} {\Lop_0}^2$ in the first term of $H_{av}$. We can therefore restrict our attention to the eigenspace of $\Lop_0$ with zero eigenvalue - i.e. the space spanned by $\ket{n^\otimes}\! := \ket{n^-}\otimes\!\ket{n^+}$, $n \geq 0$. This is natural since expressions of the form (\ref{moments}) with states from this subspace correspond to products $\mods{r_L}^{2n} \mods{r_R}^{2m}$, which are exactly those unaffected by averaging over the phase of $r_L$ and $r_R$. In section \ref{RmApp} we use the operator $U_{av}$ to reproduce Berezinskii's recursion relation for the moments of the left reflection probability, $R_m := \average{\mods{r_L}^{2m}}$.

On the zero-weight subspace, $H_{av}$ is proportional to the Casimir operator of $u(1,1)$, hence its spectrum can be obtained from the decomposition of the tensor product $T^-_0 \otimes T^+_0$ into irreducible representations. The decomposition contains the full primary series of representations, $T_\rho$ (see \cite{Vilenkin}, sections 6 and 8), each of which has an intersection with the zero-weight subspace. On $T_\rho$, the Casimir operator takes the value $-1/4 -\rho^2$ (which means all these modes decay under the action of $U_{av}$). In addition there is a state $\ket{\iota}$ in the zero-weight subspace given by,
\begin{equation}
\ket{\iota}=\sum_{n=0}^{\infty}\frac{1}{n!} \ket{n^\otimes}
\end{equation}
which corresponds to perfect reflection ($R_m\equiv 1$), and gives an identity representation of $u(1,1)$ (and so is left unaltered by $U_{av}$).

The spectral decomposition of the averaged evolution operator is apparent in many of the integral formulae for statistical properties of a disordered wire (e.g. \cite{Abrikosov} equation (15), \cite{AP} equation (64)), and is made explicit in our formalism.

\section{Relation to the Berezinskii technique}\label{Bere}

In this section we construct a representation of $u(1,1)$ such that there is a correspondence between the ``left-hand part diagrams'' Berezinskii considers \cite{Berezinskii}, and the action of the averaged evolution operator (\ref{UavEvolution}) in this representation.

The representation in question is the tensor product of the representations $T^-_{1/2}$ (representation with lowest weight $1/2$) for the retarded sector, and $T^+_{-1/2}$ (representation with highest weight $-1/2$) for the advanced sector (cf. the representations $T^-_0$ and $T^+_0$ obtained in section \ref{main}).

This representation can be realised as the orbit of the vacuum in a bosonic Fock space representation, where elements of $u(1,1)$ are quadratic in the bosonic creation and annihilation operators. The Fock space is generated by four species of boson -- corresponding to left and right moving particles (denoted by index 1 and 2, respectively), in the retarded and advanced sectors (denoted by index R and A). Creation and annihilation operators are written $\bc$ and $\ba$:
\begin{eqnarray}
\comm{\bc_{ij}}{\ba_{kl}} = -\delta_{ik}\delta_{jl}\ & , & \ i,k \in \{R,A\}\ ,\ j,l \in \{1,2\}
\end{eqnarray}
such that on the vacuum, $\ket{0_B}$,
\begin{equation}
\bc_{A1}\ket{0_B}=\ba_{A2}\ket{0_B}=\ba_{R1}\ket{0_B}=\bc_{R2}\ket{0_B}=0
\end{equation}
The action of $u(1,1)$ on this space is the sum of actions in the retarded and advanced sectors, written $\eta \to \eta^R + \eta^A$ for $\eta\in u(1,1)$, where:
\begin{eqnarray}
\mathrm{I}^i & = & (\bc_{i1}\ba_{i1} + \bc_{i2}\ba_{i2}) \quad ,\quad i\in \{R,A\} \\
\Lop_0^i & = & 1/2 (\bc_{i1}\ba_{i1} - \bc_{i2}\ba_{i2}) \\
\Lop_+^i & = & (\bc_{i1}\ba_{i2}) \\
\Lop_-^i & = & (\bc_{i2}\ba_{i1})
\end{eqnarray}

By considering the diagrams Berezinskii constructs as the timelines of retarded and advanced bosons, one can relate terms in the averaged Hamiltonian (\ref{Hav}) to the interaction vertices in the diagrams. For example, the vertex (e) in figure 2 of Berezinskii's paper\cite{Berezinskii} corresponds to the term $(\bc_{R1}\ba_{R2}\bc_{A2}\ba_{A1})$ in the averaged Hamiltonian. In general the relationship is not one-one: the term $(\bc_{R1}\ba_{R2}\bc_{R2}\ba_{R1})$ covers both vertices (b) and (c), and (b) is also given by the term $(\bc_{R1}\ba_{R1}\bc_{R1}\ba_{R1})$ (this latter degeneracy can be removed by assigning the lines in the diagrams a direction).

The correspondence between diagram vertices and terms in the averaged Hamiltonian leads to a correspondence between the left hand part diagrams and terms in the time ordered expansion of the averaged evolution operator applied to the vacuum,
\begin{equation}
U_{av}(\tau,-\infty)\ket{0_B}
\end{equation}
Further, the diagrams pertaining to $R_m$ are those which have $2m$ retarded and $2m$ advanced bosons at time $\tau$, which corresponds to a final state,
\begin{equation}
\left( \bc_{R2}\ba_{R1}\bc_{A1}\ba_{A2} \right)^m \ket{0_B} = \left(\Lop_+^R \right)^m \left(\Lop_-^A \right)^m \ket{0_B}
\end{equation}
Compare this to the matrix element for $R_m$ in the representation $T^-_0\otimes T^+_0$ from section \ref{main}:
\begin{equation}\label{Rm}
R_m = \bra{m^\otimes} U_{av}(\tau,-\infty) \ket{0^\otimes}
\end{equation}
where,
\begin{equation}
\ket{m^\otimes} := \left( \Lop_-^m \ket{0^+}\right) \otimes \left( \Lop_+^m \ket{0^-}\right)
\end{equation}

However, $R_m$ is not simply given by a matrix element of $U_{av}$ in the bosonic representation $T^-_{1/2}\otimes T^+_{-1/2}$. This is because some terms in the expansion of $U_{av}(\tau,-\infty)\ket{0}$ correspond to diagrams containing loops, which must be discarded in Berezinskii's technique. We have shown that discarding these loop-containing diagrams is equivalent to a slight modification of the representation of $u(1,1)$, from $T^-_{1/2}\otimes T^+_{-1/2}$ to $T^-_0\otimes T^+_0$.

\subsection{Recursion relation for $R_m$}\label{RmApp}

As a demonstration of the use of the construction of section \ref{construct}, we now derive Berezinskii's equations for the left-reflection amplitudes\cite{Berezinskii}. The equations relate the moments $R_m = \average{\mods{r_L}^{2m}}$ for consecutive $m$'s.

To reproduce Berezinskii's equation we need to account for the fact that he considers the evolution of the retarded and advanced Green's functions at different energies - the retarded Greens function is at an energy $\omega$ greater than the advanced Greens function. To do this we add a term $2\rmi\omega\Lop^{(-)}_0$ to the averaged Hamiltonian $H_{av}$, the superscript $(-)$ signifies that this $\Lop_0$ operator acts only on the states in $T^-_0$ - as this shift in energy only occurs in the retarded sector - and is \emph{not} the action of $\Lop_0$ in the tensor-product representation $T^-_0\otimes T^+_0$. Since the disorder statistics are position-independent and the calculation is for an infinite wire, $R_m$ will be independent of $\tau$, and so we have,
\begin{eqnarray}
0 & = & \frac{\rmd R_m}{\rmd \tau} \\
	& = & \frac{\rmd}{\rmd \tau} \bra{m^\otimes} U_{av}(\tau,-\infty) \ket{0^\otimes} \\
	& = & \sum_{n=0}^{\infty} \frac{1}{n!^2} \bra{m^\otimes} H_{av} \ket{n^\otimes}\bra{n^\otimes}
U_{av}(\tau,-\infty) \ket{0^\otimes}\\
	& = & 2\rmi\omega m R_m - \frac{1}{2l_b} m^2(R_{m+1} - 2R_{m} + R_{m-1})
\end{eqnarray}
which is the required result (c.f. \cite{Berezinskii} equation (24)).

\section{Relation to the supersymmetry technique}\label{Susy}

A supersymmetric Fock space for a 1-dimensional wire has been developed in a series of papers\cite{BF,Bocquet,BM,BFZ}. Arbitrary combinations of Greens functions have expressions which are linear in matrix elements of the evolution operator acting on the Fock space, so the supersymmetric treatment is appropriate for any set of statistical quantities, though it is not minimal even for the full set. Below we quickly outline the construction of the supersymmetric Fock space (For a more detailed exposition see \cite{Bocquet}), and show that the space we constructed in section \ref{construct} is isomorphic to a subspace of it (as a $u(1,1)$-module).

We start with a supersymmetric operator algebra with four species of bosonic operators (as for the Fock space constructed in section \ref{Bere}) and corresponding four species of fermionic operators:
\begin{eqnarray}
\comm{\bc_{ij}}{\ba_{kl}} = -\delta_{ik}\delta_{jl}\ &, &\ i,k \in \{R,A\}\ ,\ j,l \in \{1,2\}\\
\acom{\fc_{ij}}{\fa_{kl}} = +\delta_{ik}\delta_{jl} & &
\end{eqnarray}
with all other (super-) commutators zero.

The Fock space, with vacuum $\ket{0_S}$, is generated by this operator algebra and the conditions,
\begin{equation}
\bc_{A1}\ket{0_S}=\ba_{A2}\ket{0_S}=\ba_{R1}\ket{0_S}=\bc_{R2}\ket{0_S}=0
\end{equation}
\begin{equation}
\fc_{A1}\ket{0_S}=\fa_{A2}\ket{0_S}=\fa_{R1}\ket{0_S}=\fc_{R2}\ket{0_S}=0
\end{equation}

The Hamiltonian is:
\begin{equation}\label{quantH}
\mathrm{H}= \bc_{ij} (\rmi\sigma_3 V)_{jl} \ba_{il} + \rmi \omega_i \bc_{ij}\sigma_{3(jl)}\ba_{il}
+ (\ba \to \fa)
\end{equation}
where summation over repeated indices is implied. The quantities $\omega_R$ and $\omega_A$ have $\mathfrak{Im} (\omega_R)>0$, $\mathfrak{Im} (\omega_A)<0$ to ensure convergence of the expressions for the Greens functions in the Fock space. The presence of fermions makes the Hamiltonian non-Hermitian.

As has already been noted, $i\sigma_3 V$ lies in the $u(1,1)$ algebra, so (ignoring the $\omega$-term) equation (\ref{quantH}) defines an action of $u(1,1)$, which lifts to a representation of $U(1,1)$ on the Fock space. Indeed, there is a natural action of the Lie super-algebra $gl(4,4)$ on the Fock space, of which $u(1,1)$ is a sub-algebra. This structure is explained in detail in \cite{Zirnbauer96}.

The action of $u(1,1)$ on the Fock space is the sum of actions on the retarded and advanced sectors (as for the Bosonic Fock space representation in section (\ref{Bere})). On each sector the actions of the operators I, $\Lop_0$, $\Lop_-$, $\Lop_+$, are given by,
\begin{eqnarray}
\mathrm{I}^i & = & (\bc_{i1}\ba_{i1} + \bc_{i2}\ba_{i2} + \fc_{i1}\fa_{i1} + \fc_{i1}\fa_{i1}) \\
\Lop_0^i & = & 1/2 (\bc_{i1}\ba_{i1} - \bc_{i2}\ba_{i2} + \fc_{i1}\fa_{i1} - \fc_{i1}\fa_{i1}) \\
\Lop_+^i & = & (\bc_{i1}\ba_{i2} + \fc_{i1}\fa_{i2}) \\
\Lop_-^i & = & (\bc_{i2}\ba_{i1} + \fc_{i2}\fa_{i1})
\end{eqnarray}
for $i \in \{R,A\}$. Considering the retarded-sector action on the vacuum, one finds that
\begin{equation}
\Lop_0^R \ket{0_S} = \Lop_-^R \ket{0_S} = \mathrm{I}^R \ket{0_S} = 0
\end{equation}
\begin{equation}
\Lop_+^R\ket{0_S} = (\bc_{R1}\ba_{R2} + \fc_{R1}\fa_{R2})\ket{0_S} \neq 0
\end{equation}
which are precisely the conditions defining the representation $T_0^-$ above. Hence $T_0^-$ sits in the supersymmetric Fock space as the orbit of the vacuum in the retarded sector. Similarly one finds that $T_0^+$ is the orbit of the vacuum in the advanced sector. Therefore the minimal space we constructed for the local properties of the wire is isomorphic to a subspace of the supersymmetric Fock space, as representations of the dynamical algebra $u(1,1)$.

The Fock space used in section \ref{Bere} is the bosonic subspace of the supersymmetric Fock space. There the orbit of the vacuum under $u(1,1)$ turns out to be the representation $T^-_{1/2}\otimes T^+_{-1/2}$. The presence of fermions in the supersymmetric Fock space slightly alters the representation to $T^-_0\otimes T^+_0$, which is the one shown in section \ref{main} to be appropriate for the local statistical properties of the wire. The rest of the supersymmetric Fock space is unnecessary for calculations of these properties.

The observation that only a small sector of the Fock space can be reached from the vacuum under the action of the Hamiltonian was observed in \cite{BF}, and later used in \cite{Bocquet, BM} to simplify the calculation of local statistical properties of the one-dimensional Dirac Hamiltonian using the supersymmetry technique.

We have shown how to construct the space $T^-_0\otimes T^+_0$ directly as a representation of the dynamical algebra, bypassing the use of supersymmetry. We have also related this construction to the Berezinskii technique.

\end{document}